\documentclass[11pt]{article}
\usepackage[a4paper,margin=2.2cm]{geometry}
\usepackage{amsmath,amssymb,bm}
\usepackage{graphicx}
\usepackage[T1]{fontenc}
\input{glyphtounicode}\pdfgentounicode=1
\usepackage{mathptmx}
\usepackage[numbers,super,sort&compress]{natbib}
\usepackage{xcolor}
\usepackage[colorlinks=true,linkcolor=black,citecolor=blue!50!black,urlcolor=blue!50!black]{hyperref}
\usepackage{caption}
\usepackage{titlesec}
\titleformat{\section}{\large\bfseries}{}{0pt}{}
\titleformat{\subsection}{\normalsize\bfseries}{}{0pt}{}
\titlespacing*{\section}{0pt}{14pt}{6pt}
\titlespacing*{\subsection}{0pt}{10pt}{3pt}
\usepackage{pdfpages}

\newcommand{\vsf}{v_{\rm sf}}
\newcommand{\urlprefix}{}
\begin{document}

\begin{center}
{\LARGE\bfseries Sound-wave interference reveals the\\[2pt] symmetry of a quantum transition\par}
\vspace{10pt}
{\large Sergei Studenikin$^{1,2,*}$, Yousef Karimi Yonjali$^{2}$, Alex Bogan$^{2}$, Andy Sachrajda$^{2}$ and Shabir Barzanjeh$^{1,*}$\par}
\vspace{4pt}
{\small $^1$Department of Physics and Astronomy, University of Calgary, Calgary, Alberta T2N~1N4, Canada\\
$^2$Quantum and Nanotechnologies Research Centre, National Research Council Canada, Ottawa, Ontario K1A~0R6, Canada\par}
\end{center}
{\let\thefootnote\relax\footnotetext{$^*$Corresponding authors. E-mail: \href{mailto:Sergei.Studenikin@nrc-cnrc.gc.ca}{Sergei.Studenikin@nrc-cnrc.gc.ca} (S.S.), \href{mailto:shabir.barzanjeh@ucalgary.ca}{shabir.barzanjeh@ucalgary.ca} (S.B.)}}

\vspace{6pt}
\begin{quote}\small
A single sound quantum emitted from two places interferes with itself, like a particle passing through two slits. The fringe position reveals whether the sources emit with the same sign (symmetric) or opposite signs (antisymmetric). Knowing this symmetry is important because qubits in solids lose energy by emitting such sound quanta, phonons. For a qubit shared between two sites, antisymmetric emission cancels at long wavelengths, suppressing energy loss, whereas symmetric emission does not. Lifetime measurements give only the total loss and cannot distinguish them. Such fringes appear in double quantum dots, but their position has not been analysed. Here we measure it for single phonons emitted as one hole tunnels between two gallium arsenide quantum dots. Using a quantitative model of sound focusing in the crystal, we find that the two dots emit antisymmetrically, as expected for a moving charge, with only a small symmetric part. Spin-flip transitions in a magnetic field weaken the fringes without shifting them. For double-dot charge qubits, antisymmetric emission is predicted to reduce phonon losses strongly at the frequencies where qubits operate, and phonon interferometry gives a direct way to measure how qubits release energy and to design longer-lived qubits.
\end{quote}

A double-slit experiment, whether with photons or electrons, produces fringes only as long as nothing records which slit each particle passed through~\cite{Tonomura1989,Scully1991}. Emission shared between two sites follows the same principle, because the emitted quantum has an amplitude to leave from each site. The fringe period gives the site separation, and the fringe position gives the relative sign, which defines the symmetry of the emitter. Such collective emission has been observed for distant quantum dots~\cite{Tiranov2023} and for superconducting atoms coupled to microwaves or sound at several points~\cite{Kannan2020,Andersson2019}, with phases set by design.

The symmetry decides how fast many qubits lose energy. Charge and spin qubits in semiconductors relax by emitting phonons, which limits their lifetimes~\cite{Kloeffel2013,Burkard2023,Petersson2010,Petit2020}, and they are often spread over two sites, as in double quantum dots. If the two sites emit with opposite signs, their waves cancel when the wavelength exceeds the site separation, so the device geometry can suppress this loss at chosen frequencies. The sign also decides which modes of an acoustic resonator a qubit can couple to, a selection rule for quantum acoustics and phonon links between qubits~\cite{Chu2017,Satzinger2018,Bienfait2019,vonLupke2022,Wollack2022,Bild2023}. Finally, it shows whether a transition moves charge between the sites or couples locally to spin, strain or valley. Relaxation measurements give only the total rate and cannot reveal the sign.

Double quantum dots (DQDs) are a natural test bed, since a charge tunnelling between them can emit its phonon at either dot~\cite{Brandes1999}. Sweeping the phonon energy produces periodic fringes in the current, seen for emitted and absorbed phonons~\cite{Fujisawa1998,Roulleau2011,Hofmann2020,Granger2012}. Earlier work analysed the fringe spacing. The absolute position, which carries the sign, has not been used, because interpreting it requires a quantitative model of sound propagation and focusing in an anisotropic crystal. Charge-coupled emission is expected to be antisymmetric, but for holes this is not guaranteed. Strong spin--orbit coupling lets the spin flip during tunnelling~\cite{Maurand2016,Hendrickx2020,Hendrickx2021,Froning2021,Scappucci2021,Bogan2017,Wang2024}, coupling of the hole spin to strain could add a symmetric component~\cite{Kloeffel2013,Scappucci2021}, and it is open whether spin-flip transitions preserve the interference between the dots~\cite{Srinivasa2013,Danon2013,Gawarecki2021,Karwat2021,Gawelczyk2021,Gawelczyk2018}.

\begin{figure}[!t]
\centering
\includegraphics[width=\textwidth]{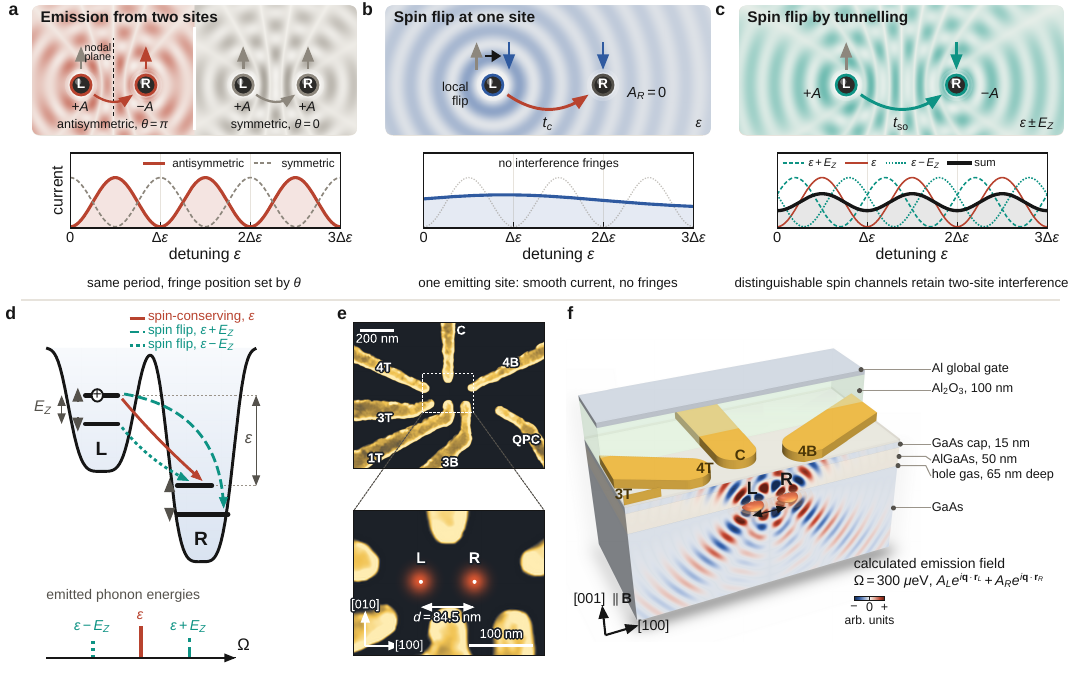}
\caption{\textbf{A two-site phonon interferometer.} \textbf{a}--\textbf{c},~Concept (theory). Top, a hole tunnels from dot L to dot R. Grey and coloured arrows show its spin before and after the transition, and the background shows the emitted phonon wave. Bottom, idealised single-mode current versus detuning $\varepsilon$, the energy difference between the dot levels, with fringe period $\Delta\varepsilon$. Spin-conserving transitions (red) emit phonons of energy $\Omega\approx\varepsilon$, and spin-flip transitions (teal) of $\Omega\approx\varepsilon\pm E_Z$, with $E_Z$ the Zeeman energy. \textbf{a},~The two dots emit with opposite signs in an antisymmetric emitter (left, $A_R=-A_L$, intrinsic phase $\theta=\pi$) and with the same sign in a symmetric one (right, $A_R=+A_L$, $\theta=0$). The dashed line marks the antisymmetric nodal plane. Below, the antisymmetric rate (red) vanishes at $\varepsilon=0$ and at multiples of $\Delta\varepsilon$, where the symmetric rate (dashed) is largest. \textbf{b},~A spin flip at one dot only would leave a single emitter ($A_R=0$) and a current (blue) without two-site fringes (grey dotted). \textbf{c},~If the spin flips during tunnelling, both dots still emit. The fringes of the channels emitting at $\varepsilon+E_Z$ (dashed) and $\varepsilon-E_Z$ (dotted) are shifted by $\mp E_Z$ from those of the spin-conserving channel (red), and their sum (black) has lower contrast at the same position. \textbf{d},~Spin-resolved levels and the resulting phonon spectrum, with colours and line styles as in \textbf{c}. \textbf{e},~False-coloured electron micrograph of the gates (top) and the central region (bottom). Gates 3T and 3B are plungers, 4T, 4B, C and 1T set the barriers, and the quantum point contact (QPC) senses the charge. The dots, not imaged, are drawn at the fitted separation $d$ and centred as electrostatic simulations suggest~\cite{Bogan2017,Bogan2021}. \textbf{f},~Cross-section along the dot axis, layers to scale, with $B$ along [001]. The colour map is the calculated phonon field at $\Omega=300~\mu$eV from amplitudes $A_L$ and $A_R$ at positions $\mathbf{r}_L$ and $\mathbf{r}_R$, with wave vector $\mathbf{q}$ (theory).}
\label{fig:1}
\end{figure}

In this work, we use a single-hole DQD in gallium arsenide (GaAs) as a two-site phonon interferometer (Fig.~\ref{fig:1}). From the absolute position of its fringes, we determine, to our knowledge for the first time, the spatial symmetry of phonon emission from a two-site transition. The fringes lie close to the antisymmetric prediction and half a period from the symmetric one. Within a tested acoustic model, this fixes the symmetry of the emitter, not the microscopic coupling behind it.

\section*{A two-site phonon interferometer}

When a hole tunnels from the left dot (L) to the right dot (R) at zero magnetic field, it emits a single phonon whose energy equals the detuning $\varepsilon$, the energy by which the level of R lies below that of L (Fig.~\ref{fig:1}a). The initial and final states are the two eigenstates of the DQD, $|a\rangle=\alpha|L\rangle+\beta|R\rangle$ and $|b\rangle=\beta|L\rangle-\alpha|R\rangle$, where $|L\rangle$ and $|R\rangle$ denote the hole in each dot, $\alpha^2+\beta^2=1$ and $\beta$ is small when the tunnel coupling is weak compared with $\varepsilon$. Both states extend over both dots, so the phonon can be emitted at either dot. The two paths leave the same phonon, with wave vector $\mathbf{q}$ of magnitude $|\mathbf{q}|\approx\varepsilon/\hbar c$ for a sound velocity $c$, so they cannot be told apart and their amplitudes add,
\begin{equation}
\mathcal{A}(\mathbf{q})=A_Le^{i\mathbf{q}\cdot\mathbf{r}_L}+A_Re^{i\mathbf{q}\cdot\mathbf{r}_R}\propto A_L+A_Re^{i\mathbf{q}\cdot\mathbf{d}},
\label{eq:amp}
\end{equation}
where $A_j$ is the emission amplitude of dot $j$, $\mathbf{r}_j$ its position and $\mathbf{d}=\mathbf{r}_R-\mathbf{r}_L$. For equivalent dots the amplitudes differ only by an intrinsic phase $\theta$, $A_R=e^{i\theta}A_L$, and the emission rate becomes
\begin{equation}
|\mathcal{A}(\mathbf{q})|^2\propto2|A_L|^2\bigl[1+\cos(\mathbf{q}\cdot\mathbf{d}+\theta)\bigr].
\label{eq:fringe}
\end{equation}
The propagation phase $\mathbf{q}\cdot\mathbf{d}$ grows with the phonon energy, because a larger $\varepsilon$ means a shorter wavelength, and equals about $\varepsilon d/\hbar c$ along the dot axis. The intrinsic phase $\theta$ is fixed by the transition and expresses the symmetry of the emitter under reflection through the midpoint between the dots. A symmetric emitter ($\theta=0$, $A_R=+A_L$) emits most strongly at long wavelengths. An antisymmetric emitter ($\theta=\pi$, $A_R=-A_L$) does not emit at long wavelengths and emits most strongly when the dots are half a wavelength apart. As $\varepsilon$ is tuned, both give fringes with the same period $\Delta\varepsilon\approx hc/d$, one additional wavelength between the dots, but their maxima are displaced by half a period. Each tunnelling event emits one phonon, which interferes only with itself~\cite{Qiao2023}, so the fringes build up in the average rate of many events.

The intrinsic phase follows from how the charge moves. The phonon couples to the transition charge density, $\rho_{ba}(\mathbf{r})=\psi_b^*(\mathbf{r})\psi_a(\mathbf{r})$, the product of the final- and initial-state wavefunctions. The two states are orthogonal, so $\rho_{ba}$ integrates to zero and the transition moves charge from one dot to the other without changing the total charge. Its contributions at the two dots have equal magnitude and opposite sign (Methods), so the phonon couples only to the difference between the lattice distortions at the two dots. The two waves start half a cycle apart, as from an oscillating acoustic dipole with a nodal plane midway between the dots.

The same picture covers transitions that also flip the spin (Fig.~\ref{fig:1}b,c). A record of which transition occurred does not destroy the interference as long as it says nothing about the emitting dot. A spin flip driven by a lattice coupling at one dot only has $A_R=0$, and its fringes vanish. If instead the spin turns while the hole tunnels, through spin--orbit coupling, the initial and final states remain orthogonal and both dots still emit with $\theta=\pi$. The spin flip changes the phonon energy to $\varepsilon\pm E_Z$, where $E_Z$ is the Zeeman energy (Fig.~\ref{fig:1}d), which identifies the transition but not the site. The spin channels keep fringes of the same period and intrinsic phase, shifted by their energy offsets, and add as intensities to a sum with lower contrast at an unchanged position.

\subsection*{Device and emission rate}
We study this emission in a single-hole DQD defined by gates on an undoped GaAs/AlGaAs heterostructure, with the dots along the [100] crystal axis and a quantum point contact sensing their charge (Fig.~\ref{fig:1}e,f). With a source--drain bias of 1.5~mV, every hole that passes through the device makes one interdot step and relaxes predominantly by emitting one phonon of energy $\Omega=\sqrt{\varepsilon^2+4t_c^2}\approx\varepsilon$, where $t_c\ll\varepsilon$ is the tunnel coupling, so a current of 10--40~pA corresponds to about $10^8$ single-phonon emissions per second. The current is then proportional to the relaxation rate of this step~\cite{Brandes1999},
\begin{equation}
\Gamma_{\rm in}(\varepsilon)=\frac{t_c^2}{\Omega^2}\,\mathcal{K}(\Omega),\qquad \mathcal{K}(\Omega)\propto\sum_\lambda\int d^2\hat{\mathbf{n}}\;\frac{q_\lambda}{c_\lambda^2}\,|M_\lambda(\hat{\mathbf{n}})|^2|F(\mathbf{q}_\lambda)|^2\bigl[1-\cos(\mathbf{q}_\lambda\cdot\mathbf{d})\bigr].
\label{eq:rate}
\end{equation}
The prefactor $t_c^2/\Omega^2=(\alpha\beta)^2$ is the squared strength of the transition charge density. The emission kernel $\mathcal{K}$ sums over the three acoustic branches $\lambda$ and all directions $\hat{\mathbf{n}}$, with wave vector $\mathbf{q}_\lambda=(\Omega/\hbar c_\lambda)\hat{\mathbf{n}}$, direction-dependent sound velocity $c_\lambda$, piezoelectric and deformation-potential matrix element $M_\lambda$ and a form factor $F$ for the extent of the emitting charge. Sound in GaAs is strongly anisotropic, so the fringes are carried by a few focused phonon rays along the dot axis, whose geometric (Maslov) phases shift the fringe pattern by a calculable amount (Methods).

\section*{Measurement of the intrinsic phase}

\begin{figure}[!t]
\centering
\includegraphics[width=\textwidth]{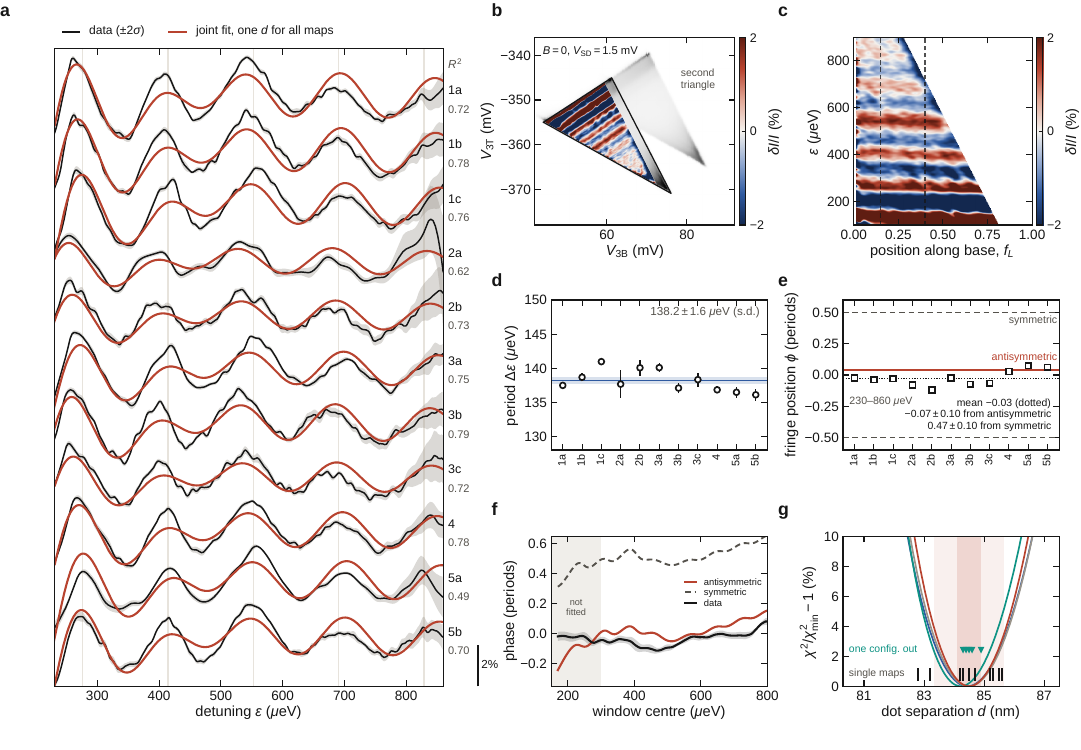}
\caption{\textbf{The fringe position identifies an antisymmetric emitter.} \textbf{a},~Relative current change $\delta I/I$ versus detuning $\varepsilon$ for the eleven maps, labelled by configuration (1--5) and repeat (a--c), offset vertically (black, $\pm2$ s.d. bands). Red curves are the joint fit (theory), with one dot separation $d$ for all maps and an amplitude and a small detuning offset for each. Numbers give the coefficient of determination $R^2$, and grey lines mark multiples of the mean period. \textbf{b},~Current at $B=0$ and source--drain bias $V_{\rm SD}=1.5$~mV versus the plunger-gate voltages $V_{\rm 3B}$ and $V_{\rm 3T}$ (configuration 1). Grey is the raw current (0--45~pA), including a second, unused triangle. Inside the analysed triangle (outline), $\delta I/I$ shows the fringes as stripes parallel to its base. Its upper edge, without a defined background, shows only the raw current. \textbf{c},~The same data versus $\varepsilon$ and the position $f_L$ along the base. Dashed lines bound the range averaged in \textbf{a}. \textbf{d},~Fringe period $\Delta\varepsilon$ of each map, which gives $d$ (circles, bootstrap errors), with the mean (line) and its standard error (band). \textbf{e},~Fringe position $\phi$ of each map, which gives the intrinsic phase, in the 230--860~$\mu$eV window, measured from the resonance without offsets (squares). Lines are predictions for antisymmetric ($\theta=\pi$, red) and symmetric ($\theta=0$, dashed) emitters along [100] (theory) and the mean of the maps (dotted). \textbf{f},~Phase in sliding 140~$\mu$eV windows versus window centre, for the data (black, mean of the eleven maps, standard error over configurations) and for noise-free antisymmetric (red) and symmetric (dashed) theory at $d=84.5$~nm, analysed identically. Grey marks windows below the fit window. \textbf{g},~Rise of $\chi^2$ above its minimum $\chi^2_{\rm min}$ versus $d$ for four effective emitter sizes with in-plane length $\ell$ and vertical length $a_z$ (red, $\ell=4.5$~nm and $a_z=5$~nm; blue, $\ell=5$~nm; teal, $\ell=4$~nm; grey, $a_z=6.5$~nm). Bands give the statistical (dark) and total (light) uncertainty of $d$. Ticks mark single-map optima and triangles leave-one-configuration-out values.}
\label{fig:4}
\end{figure}

At zero field all transitions emit phonons of the same energy, so the fringes come from a single emission channel. We recorded eleven maps at five barrier-gate settings, called configurations 1--5, with letters marking repeated maps (Supplementary Information). Dividing the current by a smooth background reveals fringes of 1--2\% as stripes parallel to the base of the bias triangle, the line of zero detuning (Fig.~\ref{fig:4}b). The stripes depend only on $\varepsilon$ (Fig.~\ref{fig:4}a,c), as expected for emission between the dots rather than lead features.

The fringe period, which gives the dot separation, is $138.2\pm1.6~\mu$eV (mean $\pm$ s.d. over maps) and varies by at most about 2\% between maps, although the barrier gates differ by up to 24~mV (Fig.~\ref{fig:4}d). We express the position as a phase $\phi$, in units of the period, with fringe maxima at $\Omega_n=(n+\phi)\Delta\varepsilon$ for integer $n$, measured from the interdot resonance. Including the geometric phase of the focused rays, the acoustic model predicts $\phi=+0.04$ for an antisymmetric and $\phi=-0.50$ for a symmetric emitter. The eleven maps give $\phi=-0.03$ in the 230--860~$\mu$eV window, with no adjustable offsets (Fig.~\ref{fig:4}e). Including the uncertainty in the zero of detuning (Supplementary Information), this lies $-0.07\pm0.10$ period from the antisymmetric prediction and $0.47\pm0.10$ period from the symmetric one, nearly the largest possible separation.

\subsection*{Robustness of the acoustic model}
Translating the fringe position into $\theta$ relies on the acoustic model. Its uncertainty is below 0.035 period for the kernels that describe the data, and varying the size of the emitting charge up to 20~nm, or the weights of the two transverse phonon branches as far as the measured period allows, moves the antisymmetric prediction by at most about 0.15 period (Supplementary Information). The measured phase is also closer to the antisymmetric value in every energy window, and within about 0.1 period of it above 300~$\mu$eV (Fig.~\ref{fig:4}f). The emitter is antisymmetric, $\theta=\pi$, and a symmetric emitter is strongly disfavoured.

A joint fit of all maps, with one dot separation and literature values for the material constants, gives $d=84.5\pm0.4_{\rm stat}\pm1.1_{\rm syst}$~nm and explains 71\% of the variance on average (Fig.~\ref{fig:4}a,g). This fit allows small detuning offsets, so it tests the period and shape of the fringes, not their position. It requires an emission cutoff at a short length, $\ell=4.5$~nm, much shorter than the expected size of the dot orbital, whose origin, like that of the absolute emission rate, remains open (Methods).

\section*{Emission in a magnetic field}

A magnetic field $B$ splits each level by the Zeeman energy $E_Z=g\mu_BB$, where $g$ is the hole $g$ factor along the field and $\mu_B$ the Bohr magneton. The resulting four states are mixed by $t_c$ and by a spin-flip tunnelling $t_{\rm so}$ from spin--orbit coupling (Fig.~\ref{fig:1}d). The emission amplitude at each dot has a main part coupled to the charge, $A_j^{(c)}=g_{\mathbf{q}}F_j\langle b|P_j|a\rangle$, where $P_j$ projects onto dot $j$, $g_{\mathbf{q}}$ is the charge coupling and $F_j$ the form factor, and a weaker part that couples the lattice directly to the spin~\cite{Srinivasa2013,Kloeffel2013}. For identical dots, $F_L=F_R\equiv F$, and since $P_L+P_R=1$, the charge parts sum to the overlap of the two states,
\begin{equation}
A_L^{(c)}+A_R^{(c)}=g_{\mathbf{q}}F\,\langle b|a\rangle=0 ,
\label{eq:orth}
\end{equation}
for every pair of eigenstates, whatever their spin, so a spin flip during tunnelling also gives an antisymmetric emitter.

The measured visibility limits any symmetric component. Writing $A_L=a+S$ and $A_R=S-a$, with an antisymmetric part $a$ and a symmetric part $S$, only direct spin--lattice coupling~\cite{Kloeffel2013,Srinivasa2013} or unequal dots can produce $S$, by equation~(\ref{eq:orth}). The ratio $s=S/a$ sets the fringe visibility $V$ and phase shift $\delta$ (Methods). The largest fitted zero-field visibility, 0.58 of the ideal value, is a lower limit on $V$, because lead limitation and unequal dots can only lower it. An $S$ in phase with $a$ then obeys $|s|\leq0.52$ and carries at most about 21\% of the intensity, and an $S$ in quadrature is limited by the measured phase to $|s|\leq0.93$ (Supplementary Information).

Spin channels with different phonon energies are distinguishable, so they add as intensities~\cite{Scully1991}, each shifted along the detuning axis by its energy offset, 0 or $\pm E_Z$. If the two spin-flip channels have equal weight and real $s$, the fringe contrast $C$ becomes
\begin{equation}
\frac{C(B)}{C(0^+)}=\frac{|1+q_{\rm sf}\cos(2\pi E_Z/\Delta\varepsilon)|}{|1+q_{\rm sf}|},\qquad q_{\rm sf}=\frac{2\widetilde V_{\rm sf}w_{\rm sf}}{\widetilde V_0w_0},
\label{eq:q}
\end{equation}
where $w_0$ and $w_{\rm sf}$ are the weights of the spin-conserving channel and of each spin-flip channel and $\widetilde V_j=V_j\cos\delta_j$ is a signed visibility, positive for an antisymmetric and negative for a symmetric emitter. The shifted patterns are most out of step near $E_Z=\Delta\varepsilon/2$ (0.85~T) and back in step near $E_Z=\Delta\varepsilon$ (1.7~T), while the fringe maxima stay in place.

\begin{figure}[!t]
\centering
\includegraphics[width=\textwidth]{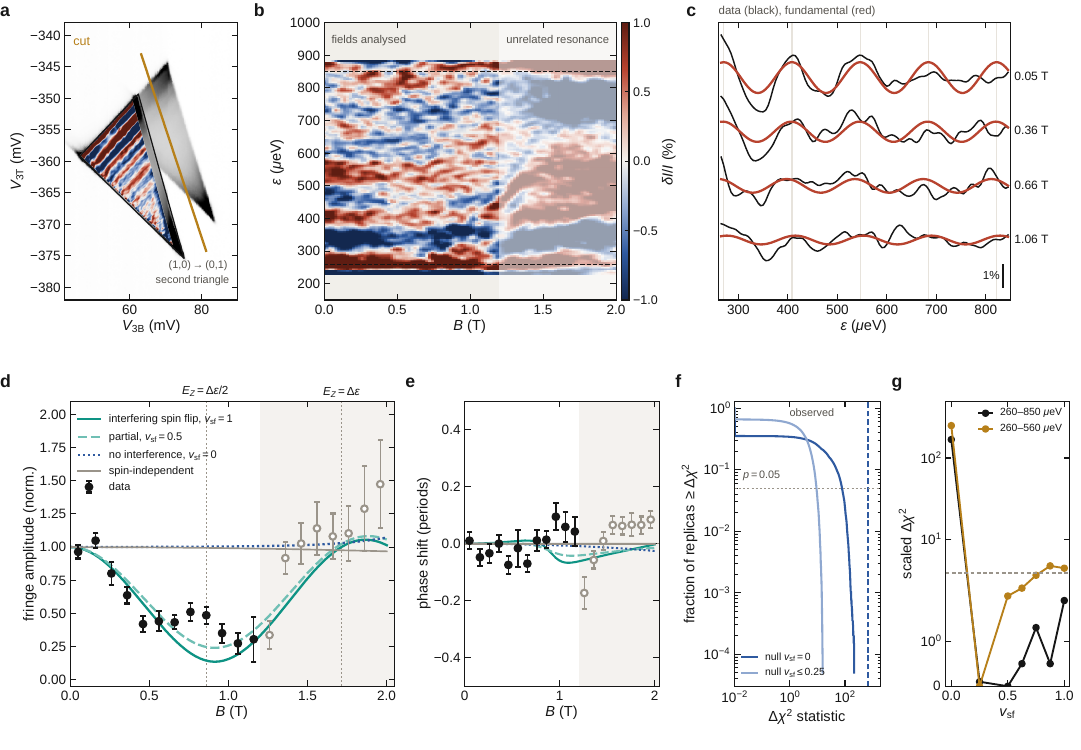}
\caption{\textbf{A magnetic field lowers the contrast at a fixed fringe position.} \textbf{a},~Zero-field map of configuration 3, shown as in Fig.~\ref{fig:4}b, with the raw current in grey (0--18~pA) and $\delta I/I$ ($\pm2\%$ colour scale) inside the first triangle. The field sweep follows the cut (orange) through the second triangle, where the hole moves from $(1,0)$ to $(0,1)$, with $(N_L,N_R)$ the hole numbers on the two dots. \textbf{b},~Measured $\delta I/I$ along the cut versus field $B$ and detuning $\varepsilon$. The fringes fade but stay in place. Dashed lines bound the analysis window, and above 1.2~T an unrelated resonance crosses it. \textbf{c},~Measured profiles at four fields (black) with fitted sinusoids of period $\Delta\varepsilon$ (red), offset vertically. Grey lines mark the fringe maxima of the lowest-field profile. \textbf{d},~Fringe amplitude relative to $B=0$ (circles, data in 0.1~T bins), with vertical lines at $E_Z=\Delta\varepsilon/2$ and $E_Z=\Delta\varepsilon$. Lines are noise-free four-level transport predictions, not fitted to these points, for spin-flip emission from both dots with visibility $\vsf=1$ and $r=t_{\rm so}/t_c=0.8$ (solid) or $\vsf=0.5$ and $r=1.0$ (dashed), for spin-flip emission without interference ($\vsf=0$, dotted) and for spin-independent emission (grey). Only interfering spin-flip emission reproduces the loss of contrast. Open grey circles, above 1.2~T, are excluded because of the unrelated resonance. \textbf{e},~Fringe position relative to $B=0$ (circles, data) with the same theory lines. \textbf{f},~Fraction of 20\,000 simulated measurements with the real noise, for $\vsf=0$ and $\vsf\leq0.25$, whose misfit $\Delta\chi^2$ exceeds a given value (narrow window), with the measured value (dashed) and $p=0.05$ (dotted). \textbf{g},~Misfit $\Delta\chi^2$ versus $\vsf$ for nominal lead parameters, minimised over $r$, in the wide (260--850~$\mu$eV, black) and narrow (260--560~$\mu$eV, orange) windows (Methods). The axis is linear below 2 and logarithmic above, and the dashed line marks $\Delta\chi^2=3.84$ (95\%).}
\label{fig:2}
\end{figure}

We recorded the fringes from 0 to 2~T along one detuning cut in configuration 3 (Fig.~\ref{fig:2}a). The fringe amplitude falls to $0.42\pm0.06$ of its zero-field value at 0.46~T and stays between 0.27 and 0.51 up to 1.2~T, while the fringe position changes by less than 0.1 period (Fig.~\ref{fig:2}b--e). Above 1.2~T an unrelated resonance crosses the analysis window, and we exclude these fields.

 The field also changes the level occupations and tunnelling rates, so we compare the data with a transport model of the four levels, processed like the data, in which $t_c$ and the lead rates are fitted (Methods). Its key parameters are the spin-flip ratio $r=t_{\rm so}/t_c$ and the visibility $\vsf$ of the spin-flip fringes, which is 1 when both dots emit coherently and 0 without interference. With $\vsf=0$ the model predicts no field-dependent loss of contrast (Fig.~\ref{fig:2}d) and is rejected with $p<3\times10^{-4}$ for nominal lead rates (Fig.~\ref{fig:2}f,g). For much wider lead-rate ranges this rejection holds in the narrow detuning window but not in the wide one (Methods). Full field--detuning maps, though not an independent test, agree (Fig.~\ref{fig:3}a--c). Within the tested transport models, spin-flip transitions emit coherently from both dots.

\section*{Origin of the contrast loss}

\begin{figure}[!t]
\centering
\includegraphics[width=\textwidth]{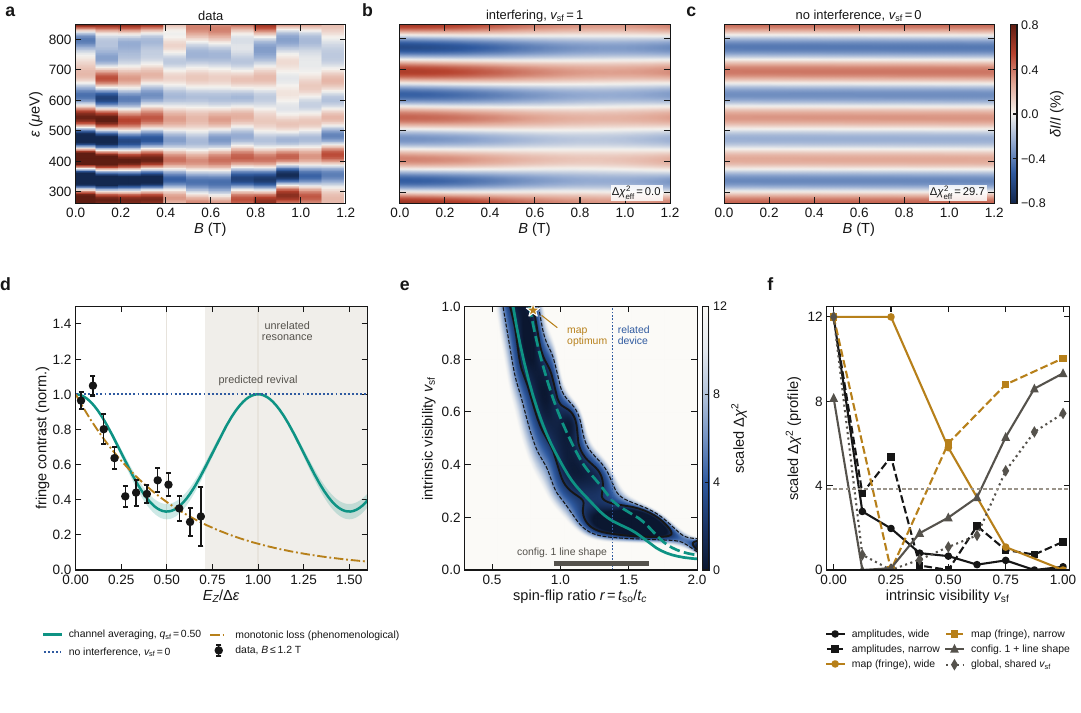}
\caption{\textbf{Contrast loss points to spin-flip emission from both dots.} \textbf{a}--\textbf{c},~Fringe component of $\delta I/I$ along the cut, filtered around the fringe period, for the data in 0.1~T bins (\textbf{a}) and for the best transport models with (\textbf{b}) and without (\textbf{c}) interference of spin-flip emission. The model maps are noise-free and shown at full resolution, and $\Delta\chi^2_{\rm eff}$ gives their misfit relative to the best model. Only the interfering model reproduces the fading of the fringes. \textbf{d},~Measured fringe contrast (circles, $B\leq1.2$~T) versus $E_Z/\Delta\varepsilon$, the Zeeman energy in units of the fringe period, with theory lines. The teal line and its 68\% band show channel averaging, equation~(\ref{eq:q}), with spin-flip weight $q_{\rm sf}=0.50$, a guide rather than a quantitative fit. The dotted line is spin-flip emission without interference ($\vsf=0$) and the dash-dotted line a phenomenological monotonic loss. Channel averaging predicts a revival at $E_Z=\Delta\varepsilon$, in the range excluded because of the unrelated resonance (shaded). \textbf{e},~Misfit $\Delta\chi^2$ of the transport model to the measured amplitudes and phases (260--850~$\mu$eV) versus the spin-flip ratio $r$ and the visibility $\vsf$, minimised over the remaining transport parameters. Along the valley, a lower visibility is offset by a larger $r$. Black contours mark approximate 68\% (solid) and 95\% (dashed) regions, not calibrated by simulation, and teal lines show constant $q_{\rm sf}=0.5$ (solid) and 0.7 (dashed). The bar marks the $r$ favoured by configuration 1, the dotted line the value $(T_N/T_F)^{1/2}$ from the measured spin-conserving and spin-flip relaxation times $T_N$ and $T_F$ of a related device~\cite{Bogan2019}, and the star the best map comparison. \textbf{f},~Misfit versus $\vsf$, minimised over all other parameters, for the amplitude fits and the map comparison in the wide (260--850~$\mu$eV) and narrow (260--560~$\mu$eV) windows, for configuration 1 and for a joint fit with one $\vsf$. All analyses disfavour $\vsf=0$. Values above 12 are clipped, and the dashed line marks $\Delta\chi^2=3.84$.}
\label{fig:3}
\end{figure}

Two effects can lower the contrast without moving the fringes, averaging over shifted spin channels or a partial loss of coherence between the dots. Channel averaging, equation~(\ref{eq:q}), reproduces the decline with $q_{\rm sf}=0.50$--0.70 (Fig.~\ref{fig:3}d), but only qualitatively, because the real channel weights are unequal and depend on energy (Supplementary Information). In this picture, antisymmetric spin-flip fringes start in step with the spin-conserving ones, giving $q_{\rm sf}>0$ and a contrast that falls as the field pulls the patterns apart, whereas symmetric ones would give $q_{\rm sf}<0$ and a contrast that first rises. In a bootstrap, $q_{\rm sf}$ is positive by at least six standard deviations, and in every detuning sub-window, as equation~(\ref{eq:orth}) predicts.

The constant-weight model captures the trend but not its details ($\chi^2\approx100$ for 23 degrees of freedom), so this sign holds only within channel averaging and reflects the spatial symmetry of emission, not the mechanism that flips the spin. Below 1.2~T a monotonic loss of contrast fits as well or slightly better. The two pictures differ near 1.7~T, where channel averaging predicts a revival to 0.9--1.0 and monotonic loss about 0.1 (Fig.~\ref{fig:3}d). The spin-flip visibility lies between about one quarter and one, because strong spin-flip tunnelling with low visibility resembles weak tunnelling with high visibility (Fig.~\ref{fig:3}e,f).

\section*{Discussion and outlook}

A single particle moving between orthogonal states cannot change its total charge, so antisymmetric emission is expected for any transition that couples to the charge of identical sites. In our hole DQD, charge-coupled emission dominates despite the strong spin--orbit coupling of heavy holes. A pure spin flip behaves differently. In an orbital shared equally by two equivalent dots it moves no charge, and identical spin--strain couplings at the two dots should make its emission symmetric, with fringes half a period away from those of charge transfer.

As in the subradiance of two emitters~\cite{Tiranov2023,Rui2020}, the antisymmetry suppresses phonon relaxation of delocalised qubits. For a charge qubit~\cite{Petersson2010,Burkard2020} with our geometry, the acoustic model extrapolated to 2--10~GHz predicts a rate of 2--40\% of that for uncorrelated dots (Supplementary Information), and delocalised spin qubits whose relaxation involves charge motion should benefit similarly~\cite{Kloeffel2013,Burkard2023,Scappucci2021,Jirovec2021,Hendrickx2024,Mi2018}. In the acoustic model, a [110] dot axis would shift the fringes by about 0.4 period, so dot spacing and orientation set where emission is suppressed.

Fields near 1.7~T, where channel averaging predicts a revival, would test whether spin-flip emission stays coherent between the dots. Spin-blockade transitions of two holes would test whether spin--orbit and spin--strain coupling add a symmetric part. Beyond quantum dots, the fringes give the site separation and the transition symmetry wherever an excitation is shared between sites, as in colour-centre pairs, coupled molecules and atoms emitting into a common mode~\cite{Tiranov2023,Kannan2020,Lange2024,Srakaew2023}.

\renewcommand{\refname}{References}

\section*{Methods}

\subsection*{Device and measurement}

The heterostructure consists, from the surface down, of a 15~nm GaAs cap, a 50~nm Al$_{0.5}$Ga$_{0.5}$As barrier and a 1000~nm GaAs layer on a buffer superlattice of 300 periods of 10~nm Al$_{0.5}$Ga$_{0.5}$As and 3~nm GaAs~\cite{Tracy2014,Ducatel2021,Studenikin2025}. The hole gas forms at the AlGaAs/GaAs interface 65~nm below the surface. Ti/Au gates define the double quantum dot (DQD) and are separated by 100~nm of Al$_2$O$_3$ from an Al global gate held at $-6.6$~V. Gates 3T and 3B, at voltages $V_{\rm 3T}$ and $V_{\rm 3B}$, act as plungers, gates 4T, 4B, 1T and C set the tunnel barriers, and a quantum point contact senses the charge. Charge states are written $(N_L,N_R)$, the numbers of holes in the left and right dots. We measure the DQD current at a source--drain bias $V_{\rm SD}=1.5$~mV with a transimpedance amplifier of gain $10^9$~V/A. The leads have an electron temperature of 102~mK, measured in the same set-up~\cite{Ducatel2021}. Each zero-field map contains $401\times201$ points, takes about eight hours and has a current noise of about 50~fA per point.

The data set comprises twelve zero-field maps at six gate configurations, full maps at 1, 2, 3 and 4~T in configuration 1, a sweep along one cut from 0 to 2~T in steps of 0.04~T in configuration 1, and the sweep of Fig.~\ref{fig:2} in configuration 3 (Supplementary Information). We do not fit configuration 6, whose fringes are much weaker than those of the other maps. An earlier tuning of a device of the same design does not resolve the phonon fringes (Supplementary Information). Relaxation data come from a published measurement on a related device~\cite{Bogan2019}, whose raw data give 17 distinct spin relaxation times within one dot. The interdot relaxation times of spin-conserving and spin-flip transfer, $T_N=3.6~\mu$s and $T_F=1.9~\mu$s, are taken from the three-level fit reported there. Predictions use the kernel with $d=84.5$~nm, $\ell=4.5$~nm, $a_z=5$~nm and a [100] axis. Supplementary Information estimates the tunnel couplings of our configurations from the absolute current.

\subsection*{Data processing}

We first place every map on a common detuning axis and then extract its relative modulation, the fractional deviation of the current from a smooth background. After subtracting the amplifier offset column by column, we locate the three edges of the first bias triangle by fitting straight lines to the maxima of the current gradient. The base and the apex then define the detuning, $\varepsilon=eV_{\rm SD}(\mathbf{V}-\mathbf{A})\cdot\hat{\mathbf{u}}/(\mathbf{C}-\mathbf{A})\cdot\hat{\mathbf{u}}$, where $\mathbf{V}=(V_{\rm 3B},V_{\rm 3T})$ is a point in the plane of the plunger voltages, $\mathbf{A}$ and $\mathbf{B}$ are the ends of the base, $\mathbf{C}$ is the apex and $\hat{\mathbf{u}}$ is the unit normal to the base. The position $f_L$ runs along the base from 0 at $\mathbf{A}$ to 1 at $\mathbf{B}$. We average profiles over $0.15<f_L<0.4$ and measure detuning from the maximum of the interdot resonance. The relative modulation is $R(\varepsilon)=I/(G_{50}*I)-1$, where $G_{50}$ is a Gaussian of 50~$\mu$eV standard deviation. Its uncertainty comes from bootstrap resampling of the columns in $f_L$. The period of each map comes from a Fourier transform between 120 and 900~$\mu$eV, and its phase from a least-squares fit of a sinusoid with this period. Resampling blocks of ten neighbouring columns gives period errors of 0.6--2.6~$\mu$eV, comparable with the scatter between maps. For the phase we quote the standard error over the five configurations, 0.03 period, because repeated maps of one configuration are not independent.

\subsection*{Emission vertex}

For a hole--phonon coupling that acts locally at each dot, the amplitude for a transition $a\to b$ between two eigenstates of the four-level Hamiltonian is
\begin{equation}
M_{ba}(\mathbf{q})=\sum_{j=L,R}e^{i\mathbf{q}\cdot\mathbf{r}_j}\,\langle b|\,P_j\,\bigl(g_{\mathbf{q}}F_j(\mathbf{q})+\bm{\sigma}\cdot\mathbf{h}_j(\mathbf{q})\bigr)\,|a\rangle\equiv A_Le^{i\mathbf{q}\cdot\mathbf{r}_L}+A_Re^{i\mathbf{q}\cdot\mathbf{r}_R},
\label{eq:vertex}
\end{equation}
where $P_j$ projects onto dot $j$, $g_{\mathbf{q}}$ is the coupling to the charge density, $F_j$ the form factor of dot $j$, $\bm{\sigma}$ the vector of pseudospin Pauli matrices and $A_j$ the total emission amplitude of dot $j$. The second term couples the lattice directly to the spin at dot $j$ and, for heavy holes, arises from strain-induced mixing with light holes~\cite{Kloeffel2013,Scappucci2021}. For the zero-field states $|a\rangle=\alpha|L\rangle+\beta|R\rangle$ and $|b\rangle=\beta|L\rangle-\alpha|R\rangle$, and neglecting the overlap of the dot orbitals, the transition charge density is $\rho_{ba}=\alpha\beta(\rho_L-\rho_R)$, where $\rho_j$ is the hole density in dot $j$ and $\alpha\beta=t_c/\Omega$. For spinor states it is in general $\rho_{ba}(\mathbf{r})=\Psi_b^\dagger(\mathbf{r})\Psi_a(\mathbf{r})$. It is positive at L and negative at R and integrates to zero because $\langle b|a\rangle=0$, which gives $A_R=-A_L$ for the charge coupling.

\subsection*{Two-site visibility and phase}

Writing the dot amplitudes as $A_L=a+S$ and $A_R=S-a$, where $a$ is the antisymmetric part ($\theta=\pi$) and $S$ the part common to both dots ($\theta=0$), each transition produces, for a single phonon mode, fringes with visibility and phase shift
\begin{equation}
V=\frac{|1-s^2|}{1+|s|^2},\qquad \delta={\rm atan2}\bigl(2\,{\rm Im}\,s,\;1-|s|^2\bigr),\qquad s=\frac{S}{a},
\label{eq:vis}
\end{equation}
where $\delta$ is measured from the antisymmetric fringe. Supplementary Information relates $V$ to the path predictability and to the overlap of the phonon wave packets emitted from the two dots.

\subsection*{Emission kernel}

The emission kernel gives the phonon emission rate versus energy from the acoustic modes of GaAs, the hole--phonon coupling and the size of the emitting charge. The phase velocities and polarisations of the three acoustic branches follow from the Christoffel equation with the elastic constants of GaAs~\cite{Vurgaftman2001}. The piezoelectric coupling is $V^{\rm PZ}_\lambda=(2ee_{14}/\epsilon_0\epsilon_r)(\hat n_x\hat n_ye_z+\hat n_y\hat n_ze_x+\hat n_z\hat n_xe_y)$, where $e$ is the elementary charge, $e_{14}$ the piezoelectric constant, $\epsilon_0\epsilon_r$ the permittivity and $e_i$ and $\hat n_i$ the Cartesian components of the polarisation $\hat{\mathbf{e}}$ and of $\hat{\mathbf{n}}$. The heavy-hole deformation potential follows from the Bir--Pikus Hamiltonian, $V^{\rm DP}_\lambda=|\mathbf{q}|[a_v\hat{\mathbf{n}}\cdot\hat{\mathbf{e}}+\tfrac{b}{2}(\hat n_xe_x+\hat n_ye_y-2\hat n_ze_z)]$, with the hydrostatic and shear deformation potentials $a_v=1.16$~eV and $b=-2.0$~eV~\cite{Vurgaftman2001}. We add the two couplings in quadrature and neglect heavy--light-hole mixing, and Supplementary Information gives the sign convention. The form factor $F=\exp[-(q_\parallel^2\ell^2+q_z^2a_z^2)/4]$, where $q_\parallel$ and $q_z$ are the in-plane and vertical components of $\mathbf{q}$ and $a_z$ is the vertical extent of the emitting charge, cuts off emission at large wave vectors. A non-perturbative $P(E)$ calculation of multiphonon processes for the measured geometry gives a Huang--Rhys factor of 0.19 and changes the golden-rule rate smoothly, from $-7\%$ to $+13\%$ across the fit window, which the background division removes. These values hold for the nominal bulk kernel, whose absolute normalisation is unresolved (Supplementary Information).

We evaluate the kernel of equation~(\ref{eq:rate}) on a grid of directions converged to better than 1\%. We tabulate it for $\ell\in\{2,3,4,6\}$~nm, $a_z\in\{3,5,8\}$~nm, dot axes at 0, 5 and $10^\circ$ from [100] and $d$ from 76 to 97~nm, with finer steps near the optimum of 0.5~nm in $\ell$ and 1--1.5~nm in $a_z$. The best kernel, $\ell=4.5$~nm and $a_z=5$~nm, lies inside the grid rather than at its edge. With an rms radius of 8~nm or more, Gaussian, exponential and disk-shaped distributions lower the mean $R^2$ to 0.31 or less. From 10~nm upward they also predict fringe amplitudes of only 0.3--1.0\%, whereas the cleanest maps show 1.0--1.1\%, although lead limitation and partial visibility can only reduce the measured amplitude. The length $\ell$ is an effective cutoff of the emission at large wave vectors, not the radius of the dot.

\subsection*{Ray analysis}

The ray analysis explains the fringe position. It finds the stationary rays, the phonon paths whose group velocity points along the dot axis, and relates their phases to the fringes. We compute the group velocity $v_{g,i}=C_{ijkl}e_je_kn_l/\rho c$ of each branch, where $C_{ijkl}$ is the elastic tensor, $\rho$ the mass density and $c$ the phase velocity, on a grid of $360\times720$ directions. We keep the directions with $|\hat{\mathbf{v}}_g\cdot\hat{\mathbf{d}}|>0.9995$, where $\hat{\mathbf{d}}$ is the unit vector along the dot axis, and refine each cluster by minimising the tangential gradient of $\hat{\mathbf{n}}\cdot\hat{\mathbf{d}}/c_\lambda$. Finite differences give the curvature of this phase function and of the emission density, from which we obtain the signature of each ray, the order of the zero of the emission density on it and its asymptotic amplitude (Supplementary Information). The phase of each ray is set by the Maslov index of the slowness surface~\cite{Wolfe1998,Msall2020}.

For $|\mathbf{q}|d=10$--40 the ray amplitudes are not accurate, because of the form factor and nearby caustics. All fits use the full kernel for this reason, and only the ray phases are compared with it. Along [100], phonon focusing~\cite{Wolfe1998} creates four slow transverse rays whose phase is opposite to that of the axial fast transverse ray. The two families partly cancel, which also makes the fringe period shorter than that of any single ray. Changing their relative weights moves the antisymmetric prediction by at most 0.05 period while the period stays within 3\%. Within this bulk acoustic model, the measured fringe position agrees with the [100] axis of the device layout, whereas an axis along [110] would give $\phi=+0.46$.

\subsection*{Joint fit}

The joint fit determines one dot separation $d$ and one broadening $\sigma$ from all fitted maps. The model rate is convolved with a Gaussian of width $\sigma$, processed exactly like the data and fitted by minimising
\begin{equation}
\chi^2=\sum_i\Bigl[\sum_\varepsilon\frac{\bigl(\eta_iM(\varepsilon-\delta\varepsilon_{0,i};d,\sigma)-R_i(\varepsilon)\bigr)^2}{\varsigma^2}+\frac{\delta\varepsilon_{0,i}^2}{(8~\mu{\rm eV})^2}\Bigr]
\end{equation}
over $230<\varepsilon<860~\mu$eV. Here $R_i$ is the measured relative modulation of map $i$, $M$ the model modulation computed from the kernel and $\varsigma=0.2\%$ a uniform residual scale. Each map has its own visibility $\eta_i$, which accounts for the weaker relative modulation when the leads partly limit the current, and its own small detuning offset $\delta\varepsilon_{0,i}$. The fitted offsets are $-9\pm8~\mu$eV and the broadening is below 3~$\mu$eV. The mean offset, $-0.07$ period, bridges the gap between the antisymmetric prediction, $+0.04$, and the measured $-0.03$, so the fitted curves match the measured phases to within 0.02 period on average. The decay of the fringes at high energy requires a lateral form-factor length $\ell=4.5$~nm, much smaller than the expected orbital radius of 15--20~nm, which we interpret as an effective cutoff of the emission. Fixing all offsets at zero gives $d=85.8$~nm, and half of the difference is included as a systematic error. The residuals are correlated over about 33~$\mu$eV and repeated maps of one configuration are not independent, so we obtain the statistical error of $d$ by leaving out each of the five configurations in turn (jackknife). In the orientation analysis the origin is fixed at the measured resonance, so the comparison of axes does not rely on the offsets.

We also tested models in which the fringe period is set by another length of the device. A single-dot model with interference from phonons reflected at the surface fits single maps comparably only for a depth of 42--44~nm, and even then raises the $\chi^2$ of the joint fit by 32\%. At the 65~nm depth of the hole gas it raises it by 530\%. Within a scalar model, the nearly matched acoustic impedance of the Al$_2$O$_3$ dielectric limits surface reflection to $|r_s|\lesssim0.1$, and the buried interfaces reflect 3--4\% in amplitude.

\subsection*{Four-level model and transport}

The four-level model predicts the current in a magnetic field, including spin-flip transitions. Its Hamiltonian is $H=\tfrac{\varepsilon}{2}\tau_z+t_c\tau_x+t_{\rm so}\tau_y(\hat{\mathbf{m}}\cdot\bm{\sigma})+\tfrac{\mu_BB}{2}[g_L\tfrac{1+\tau_z}{2}+g_R\tfrac{1-\tau_z}{2}]\sigma_z$, where the Pauli matrices $\tau$ act on the dot index and $\bm{\sigma}$ on the heavy-hole pseudospin, $\hat{\mathbf{m}}$ is the spin--orbit axis, perpendicular to the field, $\mu_B$ is the Bohr magneton and $g_L=1.44$ and $g_R=1.34$ are the $g$ factors of the two dots unless stated otherwise. Phonon emission between eigenstates occurs at the rates $W_{ab}=|\langle a|P_L|b\rangle|^2\mathcal{K}(E_b-E_a)$, and absorption at the corresponding rates weighted by the Bose occupation. The current follows from the stationary master equation for the four single-hole states and a reservoir state, with source and drain rates $\Gamma_S=\Gamma_D=10^8$~s$^{-1}$, $t_c=9~\mu$eV and $T=102$~mK, and is processed exactly like the data. For $t_c$ between 1.5 and 9~$\mu$eV the predicted amplitudes change by at most 0.04. In the first triangle the reservoir state is $(0,0)$. In the second triangle the cycle is $(1,0)\to(0,1)\to(1,1)\to(1,0)$ with reservoir state $(1,1)$, so the source and drain rates exchange roles in the lead terms while the phonon terms are unchanged.

Exchange in $(1,1)$, of order $4t_c^2/U\approx0.2~\mu$eV with $U$ the charging energy, and spin-dependent lead tunnelling both make the injection of holes spin selective. We describe this by $\kappa$, the ratio of the injection rates of the two spin states. The predicted amplitude ratios change by at most 0.06 when $\Gamma_S/\Gamma_D$ varies between 1/4 and 4, when both rates change by a factor of three, or when $\kappa$ changes by a factor of two. For the visibility test, we label as spin flips the transitions whose energy differs from the orbital gap by more than $E_Z/2$. Their kernel is replaced by $\vsf\mathcal{K}+(1-\vsf)\bar{\mathcal{K}}$, where $\bar{\mathcal{K}}$ is the kernel without the interference factor, as for emission from one dot. Classifying transitions by their spin character instead gives the same results. Every prediction is normalised to its own amplitude at $B=10^{-3}$~T, and Supplementary Information gives the rate equations and lead models.

\subsection*{Field-resolved analysis and statistics}

Along the configuration-3 cut we calibrate the detuning in two independent ways. The spin-flip line gives $\varepsilon=(58.3~\mu{\rm eV/mV})(V_{\rm base}+0.15~{\rm mV}-V_{\rm 3T})$, where $V_{\rm base}$ is the value of $V_{\rm 3T}$ at the base of the triangle, and the triangle of the zero-field map gives 60.0~$\mu$eV/mV. The two calibrations differ by about 3\%, and both give fringe periods and phases that agree with the zero-field maps within the spread from background processing. The analysis window, 260--850~$\mu$eV, lies above the spin-flip line at all fields. We average the profiles in bins of five fields, divide them by a Gaussian-smoothed background, remove a quadratic trend within the window and fit a sinusoid of period $\Delta\varepsilon=137.9~\mu$eV. Amplitudes and phases are referred to the bins below 0.12~T. The uncertainties combine the difference between even and odd fields with the 16--84\% spread over 27 variants of background width, polynomial order and window start.

The simulation test asks how often noise alone would produce the observed statistic under the null hypothesis. In each simulated sweep, the fringe at every field is replaced by the fringe that the null hypothesis predicts, scaled with the measured zero-field coefficient. The measured residuals are then added back, either reassigned in blocks of five or fifteen fields drawn with replacement or shifted circularly as a whole. Each of 20\,000 simulated sweeps per hypothesis and noise model passes through the same extraction and is compared with the full model grid ($r$ from 0.4 to 2.0, $\vsf$ in steps of 1/8). The statistic is the difference between the best $\chi^2$ under the null hypothesis and the best $\chi^2$ overall, normalised by the reduced $\chi^2$ of the best fit.

In a second version, the statistic and the null model are minimised over the transport parameters not under test, taken from the union of a nominal and an extended grid (Supplementary Information). For the nominal null model the observed statistic is 542 in the wide and 660 in the narrow window, against largest simulated values of 87--162 and 116--216 for the three noise models. On the extended grid, the wide window allows a null model only if injection is strongly spin selective~\cite{Stano2010}, with $\kappa\gtrsim3$ above about 0.5~T. This holds even when the selectivity vanishes at zero field, as time-reversal symmetry requires, and grows smoothly with field. We regard this as unlikely but cannot rigorously exclude it. With $\kappa$ between 1/2 and 2 the wide window disfavours $\vsf=0$ by scaled $\Delta\chi^2=12$, with $p=0.006$--0.09, whereas the narrow window gives Monte Carlo estimates of $p$ of at most $4\times10^{-4}$ for every tested range.

Field-dependent spin-selective injection is modelled as $\kappa(B)=1+(\kappa_{\max}-1)f(B)$, with $f=B/1.2~{\rm T}$ or $f=1-e^{-B/B_s}$ for $B_s=0.1$ and 0.3~T and $\kappa_{\max}$ from 1/8 to 8, so that $\kappa(0)=1$. For the configuration-1 maps, fringe amplitude and phase are extracted in windows that start 90~$\mu$eV above the spin-flip line, with 81 analysis variants. For these maps we test field-independent fringes, as expected for spin-independent emission, in two ways. The first uses 20\,000 simulations with the quoted errors. The second replaces the fringes of the field maps by the zero-field fringe and resamples the raw maps, and gives $p\approx0.001$.

\subsection*{Global and full-map analyses}

The transport-free analysis fits equation~(\ref{eq:q}) to the complex amplitudes of the configuration-3 bins, with $\Delta\varepsilon=137.9~\mu$eV and $g=1.41$, between the values 1.39 and 1.43 from the two calibrations. For the global analysis, the configuration-3 amplitudes and phases are computed on a grid of $t_c\in\{4,6,9,12\}~\mu$eV, $\Gamma_S/\Gamma_D\in\{1/4,1,4\}$, $\kappa\in\{1/2,1,2\}$, $r$ from 0.3 to 3 and $\vsf$ in steps of 1/8. To constrain $\vsf$ we minimise $\chi^2$ over all other parameters at each value of $\vsf$, and for the joint constraint over all parameters except $\vsf$ and $r$. The configuration-1 fringes and the line shape of the dense sweep are weighted by the effective number of independent points in the line-shape residuals. The global fit shares $\vsf$ between the two configurations while fitting $r$ and the remaining parameters separately for each.

The full-map analysis compares the measured relative modulation $R(\varepsilon,B)$ between 260 and 850~$\mu$eV and up to 1.2~T point by point with model maps processed in the same way. The models span $t_c\in\{4,6,9,12\}~\mu$eV, the same lead parameters, $r$ from 0.4 to 2.5 and $\vsf\in\{0,1/4,1/2,3/4,1\}$, with one global scale, the detuning scale fixed by the calibration and a common offset of $10~\mu$eV set from the fields below 0.12~T, where all models coincide. The noise, 0.67\% of the current, is estimated from second differences between neighbouring fields, and $\Delta\chi^2$ is scaled by $\chi^2_{\rm min}/\nu$ and by the correlation area of the residuals, which leaves 892 effective points out of 7137. The best null models of the extended grid, compared with the map but not included in its simulation calibration, are rejected by a scaled $\Delta\chi^2$ of at least 30 in the wide and 78 in the narrow window.

The fringe-only comparison keeps only the part of each map near the fringe period. Along detuning, data and model maps are filtered in Fourier space with a Gaussian of width $0.3/\Delta\varepsilon$ centred at $1/\Delta\varepsilon$, and the noise and correlation lengths are re-estimated from the filtered maps. The statistic for $\vsf=0$ is calibrated with 4000 synthetic maps per noise model. Each consists of the best null map plus the residuals of the best overall fit, reassigned in blocks of five or fifteen fields or shifted circularly, and is filtered and fitted like the data. The comparison rejects $\vsf=0$ by scaled $\Delta\chi^2=9$--62 across variants, and the observed statistic exceeds that of every synthetic map by a factor of 2--10. When the models are fitted to the fields below or above 0.6~T, $\vsf=0$ predicts the remaining fields worse than the best model by 22--92. Supplementary Information describes how $q_{\rm sf}$ depends on the detuning window, its bootstrap covariance and a flexible model with energy-dependent lead and spin-flip rates. The flexible model reproduces $q_{\rm sf}(\varepsilon)$ with $\chi^2=1.4$ and still rejects $\vsf=0$ by scaled $\Delta\chi^2=16$ and 99 in the wide and narrow windows. For the sign test, $q_{\rm sf}$ may lie between $-10$ and 3, and the phase jumps by half a period wherever $1+q_{\rm sf}\cos(2\pi E_Z/\Delta\varepsilon)$ changes sign relative to $1+q_{\rm sf}$.

\section*{Data availability}
The raw current maps, field sweeps and relaxation data analysed in this work, together with all derived profiles and fit results, are available from the corresponding author upon reasonable request and will be deposited in a public repository with a persistent identifier upon publication.

\section*{Code availability}
The analysis code in Python, including the kernel tables, joint fits, four-level transport model and simulation tests, and a Mathematica notebook reproducing the kernel and the single-map fits, is available from the corresponding author upon reasonable request and will be deposited with the data upon publication.

\section*{Acknowledgements}
We thank L.~Tracy, J.~Reno and T.~Hargett at Sandia National Laboratories for providing the GaAs/AlGaAs double quantum dot device.

\section*{Funding}
S.B. and S.S. acknowledge support from the Killam NRC Paul Corkum Fellowship. S.B. also acknowledges support from an NSERC Discovery Grant and from the University of Calgary.

\section*{Author contributions}
S.S. and S.B. conceived the idea. S.S. and A.B. performed the experiments. S.S., S.B., A.S. and A.B. analysed the data. Y.K.Y. contributed to the theory. All authors discussed the results and contributed to writing the paper.

\section*{Competing interests}
The authors declare no competing interests.

\section*{Additional information}
The Supplementary Information is appended after the main text. Correspondence and requests for materials should be addressed to S.S. or S.B.

\clearpage
\includepdf[pages=-]{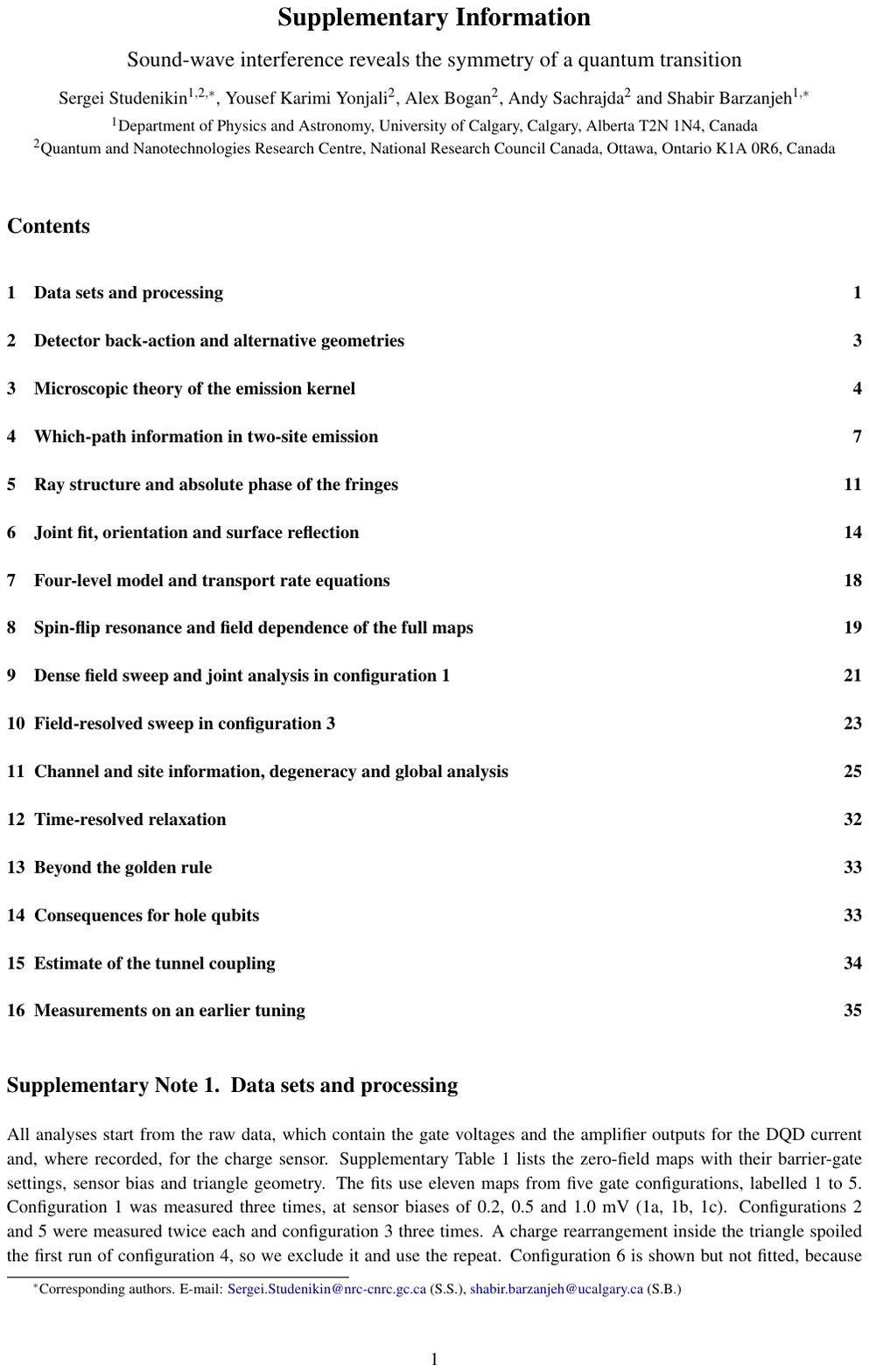}

\end{document}